# A two-dimensional position-sensitive microchannel plate detector realized with two independent one-dimensional resistive anodes

Yuezhao Zhang[1,2], Deyang Yu[1,a)], Junliang Liu[1,2], Liping Yang[3], Wei Wang[1], Xiaoxiao Li[1,2], Xiaona Zhu[1,4], Xiaoxun Song[1,5], Xinfei Hui[1,5], Wei Xi[1,5], Xin Li[1,5], Huiping Liu[1], Xiaohong Cai[1]

[1]*Institute of Modern Physics, Chinese Academy of Sciences, Lanzhou 730000, China*
[2]*University of Chinese Academy of Sciences, Beijing 100049, China*
[3]*School of Intelligent Manufacturing, Sichuan University of Arts and Sciences, Dazhou 635000, China*
[4]*School of Physics and Electronic Information Engineering, Northwest Normal University, Lanzhou 730070, China*
[5]*School of Nuclear Science and Technology, Lanzhou University, Lanzhou 730000, China*

A two-dimensional position-sensitive microchannel plate detector, with a Ø100 mm sensitive area and equipped with four integrated amplifiers, is realized with a new anode scheme. The anode is constructed by two independent one-dimensional resistive anodes, each consisting of an array of parallel copper strips connected by resistors in series on a printed circuit board. The arrays are perpendicularly aligned to realize two-dimensional position resolution, with the intervals between the adjacent strips on the PCB nearer to the MCP cut out to allow electrons passing through. The detector is tested with an $^{55}$Fe x-ray source, and a position resolution of 0.38 mm is achieved.

## I. INTRODUCTION

Microchannel plate (MCP) detectors are widely used in measuring the position of the secondary particles produced in the nuclear, atomic and molecular collision experiments.[1] A typical kind of MCP detector consists of two or three MCPs in chevron or Z-stack configuration, and an anode behind the MCPs to collect the secondary electrons.[2,3] The impact position of the incoming particles can be encoded from the output signals of the anode, which in turn makes the anode critical for the position resolution, the counting rate tolerance and other detector performances.

A type of the most commonly used anode is the resistive anode, with which the position information can be determined with the charge division method[2,3]. A two-dimensional cross-connected-pixels (CCP) resistive anode has been developed in our group.[4] Compared to the traditional resistive foil anode,[2] the CCP anode features an easy and economical manufacturing technique. It also enables position resolution without demanding an even charge allocation ratio, in contrast to the wedge-and-strip anode.[3] However, in large detecting area cases, it is hard to tune the parameters so that the anode can have even fair position resolution and at the same time a moderate counting rate tolerance, due to the large parasitic capacitance of the CCP anode. At the same time, we are currently developing a new spectrometer called momentum computed tomography (MCT) for the measurement of the momentum distribution of the secondary ions produced in ion-atom, ion-molecule or ion-solid collisions. The basic principle of the spectrometer is very similar to positron emission tomography (PET)[5] and cryogenic electron microscopy (Cryo-EM).[6] Also similar to the measuring scheme in Ref.[7] a detector with a programmed mask (aperture pattern) is required in the MCT spectrometer.

In this paper, we present a large area resistive anode, the double-layer strip-array (DLSA) resistive anode. The DLSA anode enables an order of magnitude faster response speed than the CCP anode, while still preserving its advantages of easy manufacture, charge allocation robustness and flexible geometric applicability. The design details and the electronic parameters optimization are discussed. Application of the anode in a masked detector is realized and the detector is tested with an $^{55}$Fe x-ray source.

## II. THE ANODE

### A. Anode Structure

The basic scheme of the anode is illustrated in Fig. 1. It is composed of two independent arrays of parallel strips, which are distributed on two 1.6 mm thick, 120×120 mm² PCBs of FR-4 laminate. Fig. 1(a) shows one of the strip arrays, i.e., the X-strip-array. The front side (facing the MCPs) of the PCB is composed of 85 gold-coated copper strips of width 0.6 mm and period 1.2 mm, which cover a round area of 101 mm diameter. The 0.6 mm intervals between the strips are cut out from the PCB to allow electrons to pass through. Fig. 1(b) shows the other strip array, the Y-strip-array, which covers the same area as the X-strip-array. In contrast, the gold-coated copper strips on the front side have a width of 1 mm with a period of 1.2 mm, and the 0.2 mm intervals between the strips are *not*

---

cut out. The zoomed area in each of the figures shows the details of the strips on the front side of the corresponding PCB. In practice, the two PCBs are assembled as a whole, with the X strips perpendicular to the Y strips and a gap of 1.8 mm between the two boards, as is shown in Fig. 1(c). On the back side of the anode PCBs, the adjacent strips are connected through plated vias to resistors $R_0$ implemented in an 8-pack resistor array. An end resistor $R_E^x$ or $R_E^y$ is appended to each end of the resistor series on the X or Y PCB, respectively, connecting the edge strips to the output soldering pads. This is designed to avoid drastic unbalance between the outputs when the electron cloud hits the edge strips. At last, to minimize the interactions of the anode strips with the surrounding electrodes, copper cladding is *avoided* on both sides of the anode PCBs.

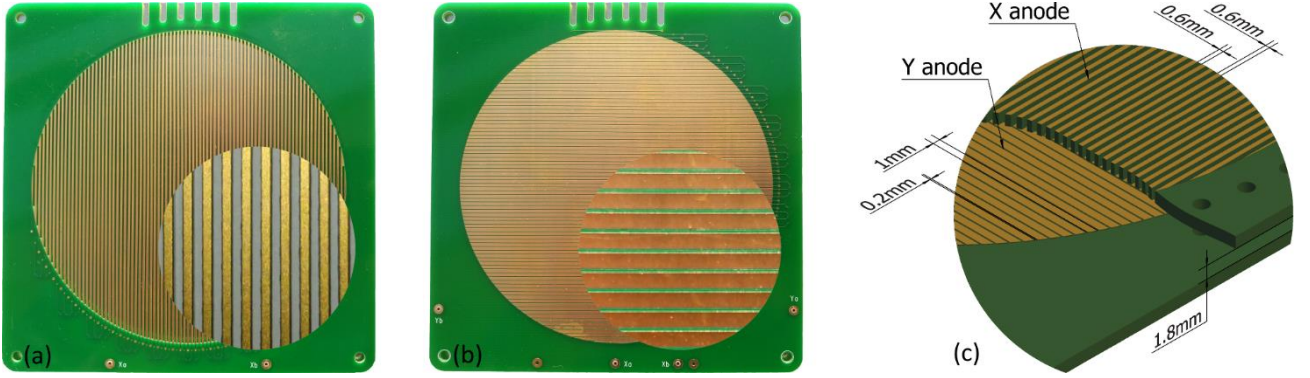

FIG. 1. Structure of the anode. (a) The anode PCB for horizontal (X) direction resolution. (b) The anode PCB for vertical (Y) direction resolution. (c) Assembly parameters of the anode. The zoomed areas in corresponding figures show the details of the strips.

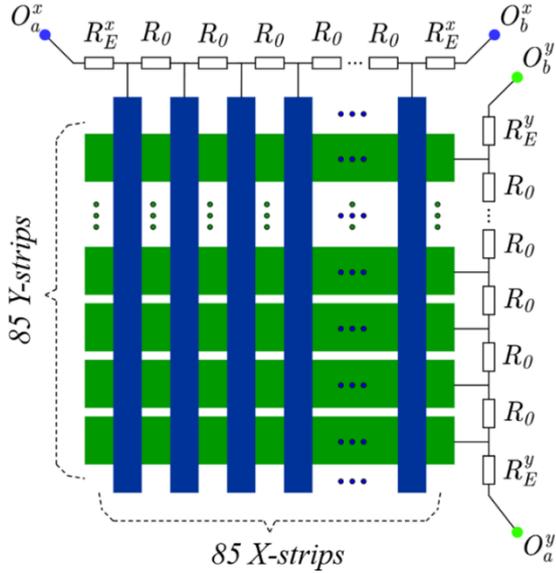

FIG. 2. Schematic illustration of the anode electronics. The X strips and Y strips are shown in blue (vertical) and green (horizontal), respectively. Denotations are $R_0$, the inter-strip resistor; $R_E^x$, the end resistor of the X-strips; $R_E^y$, the end resistor of the Y-strips; $O_a^x$ and $O_b^x$, the outputs of the X-anode; and $O_a^y$ and $O_b^y$, the outputs of the Y-anode.

## B. Working principle

The principle of the anode is sketched in Fig. 2, in which the X strips are colored in blue and the Y strips in green. When the electrons from the MCP reach the anode, part of them are received by the X strips, while the rest by the Y strips. A pair of current pulses, corresponding to the integrated charges $Q_a^x$ and $Q_b^x$, are then generated at the X anode outputs $O_a^x$ and $O_b^x$, and charges $Q_a^y$ and $Q_b^y$ at the Y anode outputs $O_a^y$ and $O_b^y$. Denoting the resistance between the electron cloud center on the anode and the outputs as $R_a^x$, $R_b^x$, $R_a^y$ and $R_b^y$ respectively for $O_a^x$, $O_b^x$, $O_a^y$ and $O_b^y$ and ignoring cross talk, the charge ratio $Q_a^x/Q_b^x = R_b^x/R_a^x$ and $Q_a^y/Q_b^y = R_b^y/R_a^y$, and the spatial coordinate of an incident particle can be calculated from the following formulae:

$$X = k_x \frac{Q_b^x}{Q_a^x + Q_b^x} + X_{offset} \quad (1)$$

$$Y = k_y \frac{Q_b^y}{Q_a^y + Q_b^y} + Y_{offset} \quad (2)$$

If the origin $(x, y) = (0,0)$ is set to the anode center, the four coefficients can be determined with the anode parameters as $k_x = -2X_{offset} = \phi R^x/R_{inter}$ and $k_y = -2Y_{offset} = \phi R^y/R_{inter}$, where $\phi = 101$ mm is the diameter of the effective area of the anode, $R_{inter}$ the sum of the $R_0$ resistors on each anode PCBs, and $R^x = R_{inter} + 2R_E^x$ and $R^y = R_{inter} + 2R_E^y$ respectively the total resistance on the X anode and the Y anode. However, the cross talk between the strips of the anode will make the charge-ratio equalities fail and the calculated spectrum be nonlinear, i.e., the calculated spatial coordinate in the spectrum will be a nonlinear function of its real value. We will return to this nonlinearity topic later in presenting the detector performance test results in section IV.

A resistive anode can be viewed as an RC transmission line with the time constant ~$RC$, where $R$ is the output resistance and $C$ is the total parasitic capacitance to ground, which is dependent on the geometric configuration of the anode and the surrounding electrodes.[8] The position resolution of an MCP detector with resistive anode depends mostly on the signal-to-noise ratio (SNR) of the subsequent spectroscopy amplifiers, which in turn depends on the anode output noise that is proportional to $C/\sqrt{R}$.[9] Though a larger $R$ is preferred to inhibit the noise, it leads to a more apparent ballistic deficit when $RC > \tau$,[9] where $\tau$ is the shaping time of the following spectroscopy amplifiers. In addition, the neighboring X and Y strips are coupled through parasitic capacitance, which leads to cross talk and consequently deteriorated spatial resolution.[10]

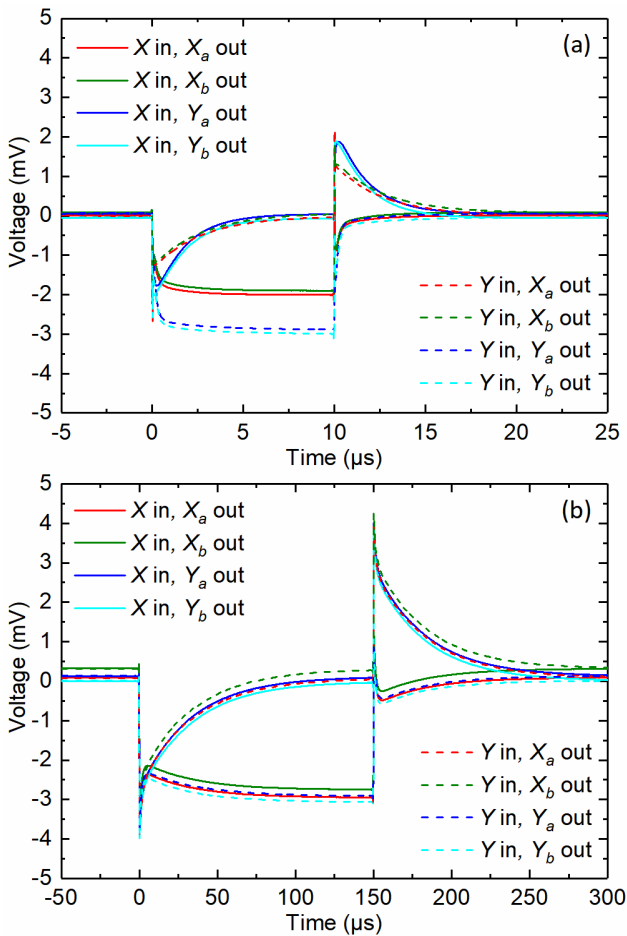

FIG. 3. Anode step response measured with a negative voltage pulse input in turn to the central x and y position of (a) the strip resistive anode and (b) the cross-connected-pixels resistive anode. The only difference between the corresponding inputs is the signal width.

### C. Electrical parameters and response speed

Considering the counting rate and position resolution requirement of the experiment, the resistors are optimized as $R_0 = 1$ k$\Omega$, $R_E^x = 82$ k$\Omega$ and $R_E^y = 42.2$ k$\Omega$, which leads to $R_E^x : R_E^y : R_{inter} \approx 0.976 : 0.502 : 1$. With these parameters, the parasitic capacitances are measured in the same electrical environment as that in the detector. The anode is carefully placed between two grounded sheet electrodes of the same area as the anode PCBs. The distance is 8 mm between the X anode and its neighboring electrode, and 6 mm between the Y anode and the other electrode. The parasitic capacitance of the anode to ground is measured to be 48 pF for the X-strip-array and 49 pF for the Y-strip-array. The parasitic capacitance between the X- and Y-strip-arrays is measured to be 47 pF, which is about two orders of magnitude smaller than that of the CCP anode of ~800 pF.[4] To further determine the response speed and optimize the spectroscopy amplifier parameters, the step response of the anode is studied in the aforementioned electrical configuration and environment.

A negative voltage pulse of amplitude 5 V, width 10 μs and rising (falling) time 5ns is firstly supplied to the middle X strip. The four outputs of the anodes are monitored with a four-channel oscilloscope with a cutoff frequency of 25 MHz. The voltage pulse is then switched to the middle Y strip, the output signals recorded with the same oscilloscope. The response curves are shown in Fig. 3(a). On the input of the pulse, the pulse-receiving-anode's outputs firstly fall fast, then gradually, to a steady negative voltage and the counterpart outputs fall almost immediately to a negative peak and then rise slowly to the baseline. An anti-symmetric phenomenon happens at the end of the input pulse. Asymmetric behavior between the X and Y outputs is observed, which is due to the asymmetry of the anode structure and the resistance. Fitting the induced waveforms with an exponential decay $e^{-t/\tau}$, the time constant $\tau$ is determined to be approximately 2.8 μs. As a comparison, the step response of the CCP anode with 81 X-pixel-strings and 80 Y-pixel-strings distributed on a PCB of the same area and connected by 1 k$\Omega$ inter-string resistors and 42.2 k$\Omega$ end resistors is measured with the same devices and arrangements, which is shown in Fig. 3(b). The decay time constant is fitted to be approximately 34.0 μs, which is more than an order of magnitude larger than that of the DLSA anode. Lowering the resistance of the CCP anode will make it respond more quickly, but as we mentioned above, a smaller anode resistance leads to a lower signal-to-noise ratio and subsequently worse position resolution.

### III. REALIZATION OF THE DETECTOR

An MCP detector is constructed with the current anode configuration, the structure of which is depicted in Fig. 4(a). The detector assembly is shielded by an aluminum alloy shell with a separate back cover. A Ø101 mm opening in the front wall allows particles to enter. A stainless steel mesh is installed on the front surface to separate the external electric field from the internal. Behind the mesh a PEEK holder with ledge is installed. A copper mask is fixed in the holder immediately behind the ledge, and a Z-stack MCP consisting of three Ø106.3 mm MCPs of 12 μm channel diameter and 55% open-area follows. The X-anode, which is about 8 mm behind the last MCP, is fixed directly to the PEEK holder. The mask and the MCPs' bias voltages are supplied by ring PCB electrodes, and the same type of ring electrodes are collocated between the last MCP and the anode to avoid distortions of the inner electric field near the edges. A high voltage PCB connected to the ring electrodes by 10 kV Kapton-insulated wire is fixed 6 mm behind the anode. This high voltage PCB is cladded with copper on both sides as much as possible, while still keeping the safety of the high voltage circuit, to shield the anode from the strong quasi-Gaussian pulses of several volts amplitude at the outputs of the spectroscopy amplifiers. The four spectroscopy amplifiers are plugged into sockets on another PCB behind the high voltage. The high voltages are supplied via an SHV coaxial connector, and the amplifiers' power supply and signal outputs are realized via LEMO coaxial connectors.

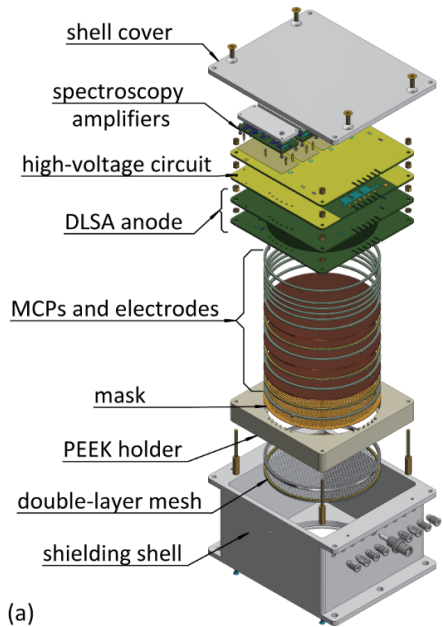
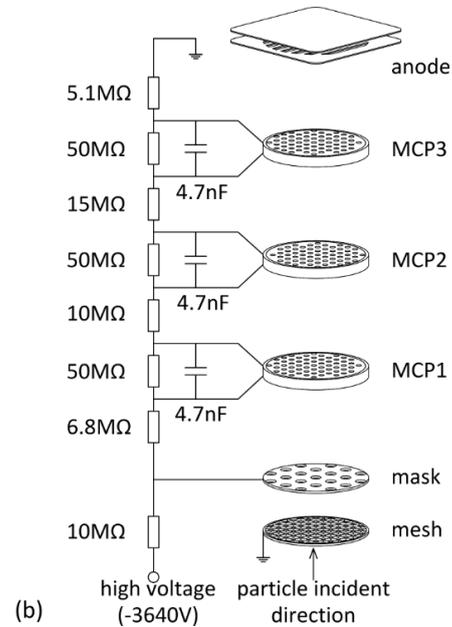

FIG. 4. Sketch of (a) the components and structure of the detector and (b) the high voltage circuit and parameters. The resistance of each MCP is about 80 MΩ.

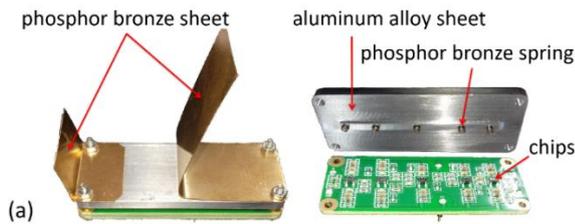
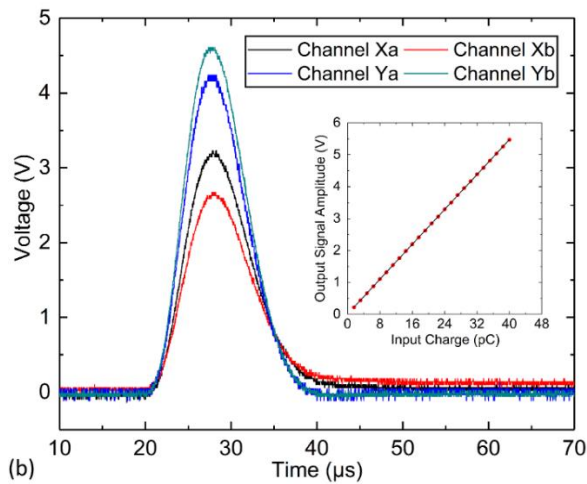

FIG. 5. The spectroscopy amplifier and its performance. (a) Photographs showing the amplifier and the assembling of the heat dissipation components. (b) Outputs of the amplifiers corresponding to a detecting event. The inset shows the linear fit to the amplifier output amplitudes at inputs ranging from 1.6 pC to 40 pC.

The high voltage circuit is shown in Fig. 4(b). A high-voltage shunt capacitor of 4.7 nF and a 50 MΩ resistor are connected in parallel with each MCP, and the measured resistance of the MCPs in parallel with their shunt resistors is about 30 MΩ. The inter-MCP resistances are a third to a half of the measured parallel MCP-shunt resistances. They are intentionally set larger than is conventional, especially the one between the middle and the last MCP, to make sure that the last MCP works in a saturated state. The resistance between the last MCP and the anode is 5.1 MΩ, corresponding to a working voltage of about 130 V in running condition. This is a balance of the following considerations: a larger voltage leads to smaller electron cloud diffusion on the anode and subsequently less edge resolution distortion of the detector, while a smaller voltage results in more even charge allocation on the anode, because less secondary electrons are generated at the X anode PCB by the bombardment of less energetic electrons.

The charge-sensitive spectroscopy amplifiers are similar to the one described in Ref.[4, 11] except that the electronic elements are more compactly integrated and that an improved grounding scheme is adopted. The shaping parameters of the spectroscopy amplifiers are determined according to the anode response time constant $\tau$, i.e., the time constant of the preamplifier is set to a comparable value of 2.5 μs and the cutoff frequency of the following main amplifier is set to 50 kHz. This leads to a full width of the output quasi-Gaussian pulse of 20 μs, which allows a counting rate of 1 kHz with a tolerance of 2% pileup. To adapt the MCPs and the following ADCs, the gain of the amplifiers are tuned to 136.9 mV/pC and the maximum output amplitude is about 5.7 V. The parameters of the four spectroscopy amplifiers are carefully adjusted to ensure consistency, and a 306 mW average power consumption of a single amplifier is observed during the detector operation. To efficiently dissipate the heat generated by the operational amplifiers, the chip top surface is connected to the detector shell consecutively by a spring, an aluminum alloy sheet and a pair of phosphor bronze sheets, as is shown in the photographs in Fig. 5(a). Fig. 5(b) displays the signals corresponding to a detecting event. A rather small noise of less than 10 mV is shown on the baseline. The linearity of the amplifiers is calibrated with a current pulse generator, with the input charge varying from 1.6 pC to 40 pC. Good linearity with a correlation coefficient of 0.999994 is shown, as is illustrated in the inset of Fig. 5(b). The amplifiers' performance is very stable during continuous work of two weeks in vacuum.

## IV. PERFORMANCE TEST

The detector is tested with an $^{55}$Fe x-ray source in a vacuum of about $2 \times 10^{-8}$ mbar. The source of diameter 10 mm is placed about 300 mm away from the detector on the detector axis. A high voltage of -3640 V is provided by an ORTEC 660 NIM module to the mask and the MCP stack, and the amplifiers' working voltage of $\pm 6$ V is supplied with an Agilent E3631A DC power supply. The amplifier outputs are transmitted via four 2 m-long 50 Ω coaxial cables to a pair of Fast ComTec 7072 DUAL ADC NIM modules that work in pulse height analyzing mode. The ADC threshold and the lower level discriminator are set to 200 mV, and the upper level discriminator is set to 6 V. The converted digital amplitudes are acquired in coincidence by a Fast ComTec MPA-3 multiparameter system. The average counting rate is about 160 Hz during the test. Any events that contain one or more saturated pulses are discarded in calculating the subsequent spectrums, which makes up 3.2% of the total counts.

### A. Charges collected by the anode

To monitor the total charge generated from the last MCP and detected by the X and Y anodes for an event, the ADC outputs corresponding to each anode are firstly summed, and the sums are summed again. The total sum spectrum is plotted in Fig. 6(a), which only takes account of the events at the central area of the two dimensional position spectrum. As can be seen, the MCP adjacent to the anode works in a good saturated state, with a peak output charge of ~100 pC. The long tail can be caused by ion feedback in a channel.[12] To estimate the signal loss due to ADC channel coincidence, an additional measurement was carried out, where the outputs of the four spectroscopy amplifiers were acquired independently. The ratio of the average ADC counting rate in the coincident measurement over that in the non-coincident measurement is larger than 0.97, i.e., few events are lost due to ADC channel coincidence. The anode sums are plotted as a two-dimensional spectrum in Fig. 6(b) with the abscissa for the X anode and the ordinate for the Y anode. Inspecting Fig. 6(b), we find that the Y anode receives more electrons than the X anode with a charge allocation ratio of $Q_y/Q_x \approx 1.5$. In optimizing the high voltage circuit parameters, we find that the charge allocation ratio can fall to one when the anode-to-MCP voltage falls to 50 V or rise to two when that voltage rises to 500 V, while the position spectrum is *not* affected evidently.

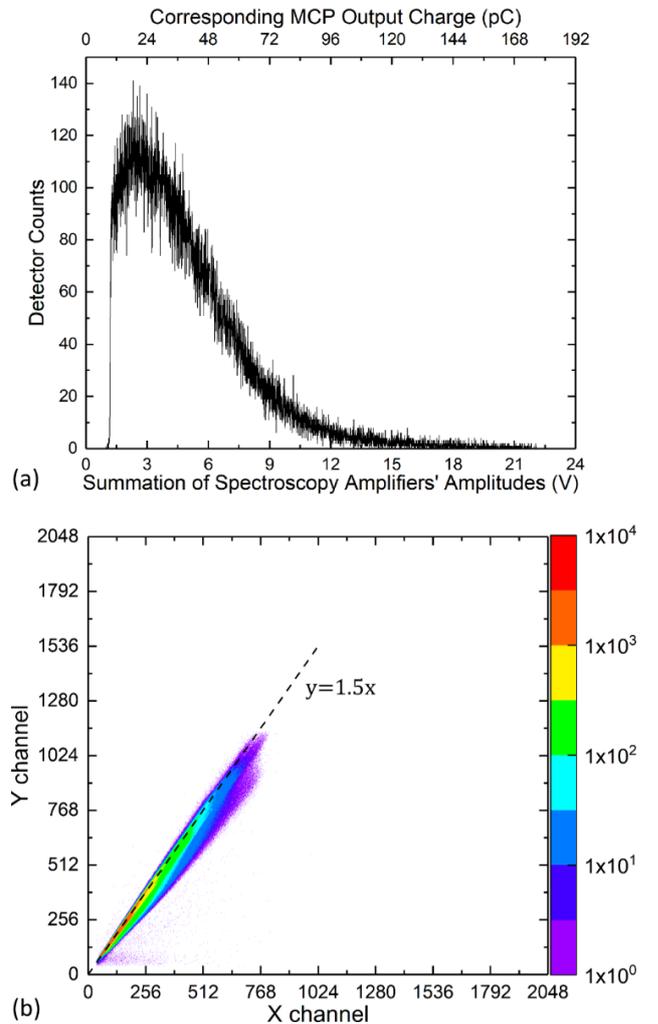

FIG. 6. (a) Distribution of the summation of the four amplifier output amplitudes, corresponding to particles detected at the central area of the two-dimensional spectrum. (b) Two-dimensional spectrum demonstrating the relationship between the X-direction and the Y-direction amplifier output amplitude sums.

This over-unity charge allocation ratio can be explained by the following picture. The electrons from the last MCP are emitted as a current pulse in the time scale of tens of nanoseconds. On approaching the X anode, they experience a non-uniform electric field that tends to drag them to the electrode and the sidewalls of the X strips, which at the same time leads to charging on the insulating sidewalls. The total charge $Q$ accumulated on an irradiated area on an anode sidewall after the electron pulse can be calculated from the charge conservation law,[13] which is the accumulation of the primarily electrons, the subtraction of the backscattered electrons and the secondary electrons, and the charge evacuation to the strip electrode during this period. Subsequently, the accumulated charges will continue to evacuate to the adjacent X anode electrode via the transport of the mobile charges and the trapped charges. For a *pre-irradiated* isolating surface, the time scale of the mobile charge evacuation can be in seconds, while that of the detrapping of the trapped charges is even larger.[14]

The charge allocation ratio of ~1.5 implies that the sum of the secondary electron yield and the backscattering coefficient should be larger than one, and comparatively the charge evacuation process during the electron pulse typically could be ignored if the anode electrode were *absent*.[13] Thus, the irradiated area would be positively charged after the electron pulse, i.e., $Q > 0$. It seems that the subsequent slow evacuation process would cause continuous positive charge accumulation on the sidewalls due to successive electron pulse bombardments, blocking the electrons of subsequent pulses from reaching the Y anode. However we didn't observe such effects in the test, neither in an experiment where the detector worked in 1 kHz counting rate for detecting a focused C$^+$ ion beam of diameter ~3 mm. This may be due to the radiation induced conductivity (RIC) effect,[15, 16] which can tremendously accelerate the charge evacuation to the X anode electrodes and thus weakens the charging effect. We hence conclude that the charge allocation ratio is dominated by the secondary electron generation and the primary electron backscattering processes.

**B. Position spectrum**

A mask as depicted in Fig. 7(a) is employed during the test. It consists of an array of round apertures, arranged in a step size of 2 mm in the X and Y directions. The central aperture diameter is 2 mm and the diameter of the other apertures is 1 mm. Substituting the values of $\phi$, $R_{inter}$, $R_E^x$ and $R_E^y$, the particle impacting position on the MCP surface is calculated from formulae (1) and (2) with $k_x = -2X_{offset} = 298.152$ mm, and $k_y = -2Y_{offset} = 202.404$ mm. An average background counting rate of 15 Hz is observed, which is likely due to radiogenic decays in the MCP itself. The background events distribute rather evenly over the detecting area except the edge zone, which is shown in Fig. 7(b). A typical position spectrum with the background subtracted is shown in Fig. 7(c), where the zoomed insets show the enlarged details of the nearby edge areas, with the exception of the lower left one showing the indicated central area of the map. Evident at $x \approx -55.9$ mm and $y \approx -38.0$ mm is a "hot spot" in the MCP due to surface contamination.

As can be seen from Fig. 7(c), the mask apertures are clearly separated in the position spectrum and the best resolution happens at the central area. However, some non-ideal defects can also be observed. Firstly, the edge ring of about 10 mm width almost loses the position information. This is due to the expansion of the electron cloud on flying from the MCP to the anode, i.e., not all of the MCP electrons are accepted by the anode strips when an event happens to be in the edge zone. Secondly, the aperture-spot shape varies across the spectrum. We find this is mainly due to the non-ideal ADC performance. Rearranging the order of the anode outputs to the ADC channels or using another pair of Fast ComTec 7072 DUAL ADC modules, we always get a different aperture-spot-shape evolution pattern. Thirdly, the spectrum exhibits certain asymmetry between the X and Y directions. The overall shape of the map exhibits an elliptical-like distribution, spanning broader in the horizontal direction. This is directly caused by the cross talk between the anode strips, which in turn is affected by the parasitic capacitance between the anode and the surrounding electrodes, the parasitic capacitance between the anode strips, and the resistance of the anode resistors. In practice, the separation nature, the structure and the resistor configuration asymmetries between the anode PCBs, together with the arrangement of the surrounding electrodes, lead to the asymmetric cross talk behaviors between the two directions and thus to the elliptical shape. For instance, we find the position spectrum is more asymmetric if the anode is nearer to the MCP, and subsequently other anode parameters should be adjusted to tune the position spectrum. Lastly, the alignment of the aperture-spot-centers in the horizontal and the vertical directions is nonlinear, which is another effect of cross talk.

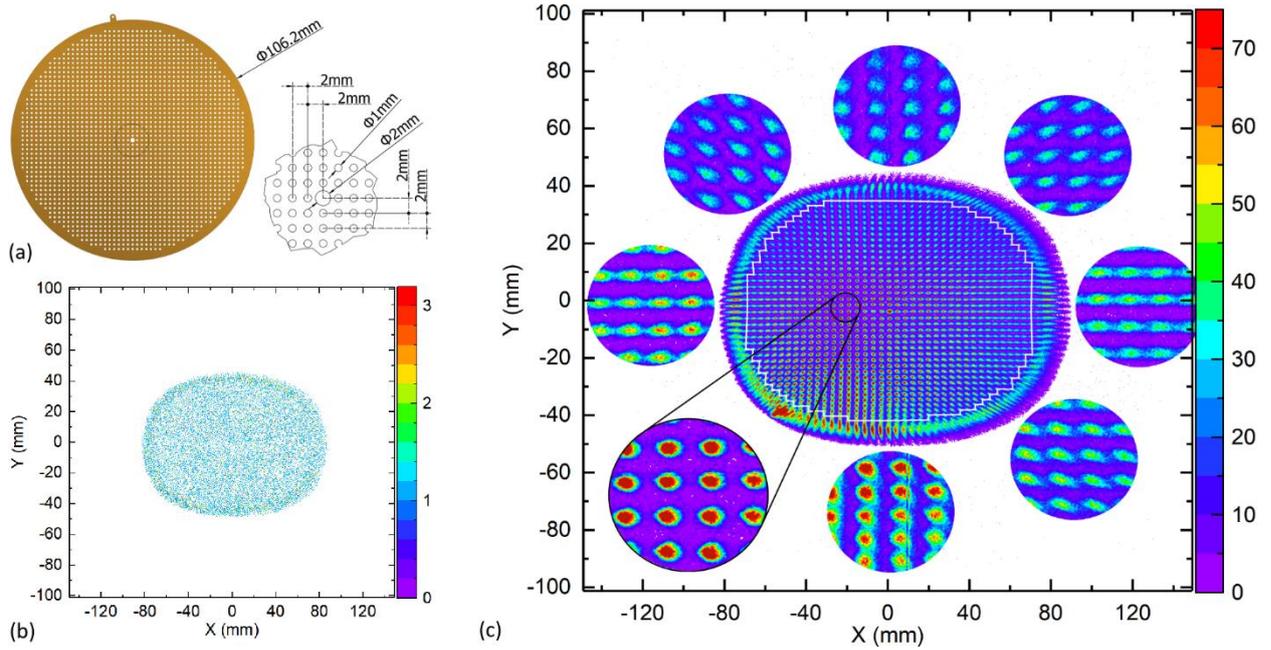

FIG. 7. (a) Detector mask used in the measurement. The inset shows the detail parameters. (b) Background spectrum measured with the radiation source removed. (c) Two-dimensional position spectrum measured with an $^{55}$Fe x-ray source illuminating the detector. The insets show enlarged areas adjacent to the outmost ring halo, with the exception of the lower-left one that corresponds to the indicated central area. The white polygon indicates the effective detecting area, on which the linearization calibration is performed.

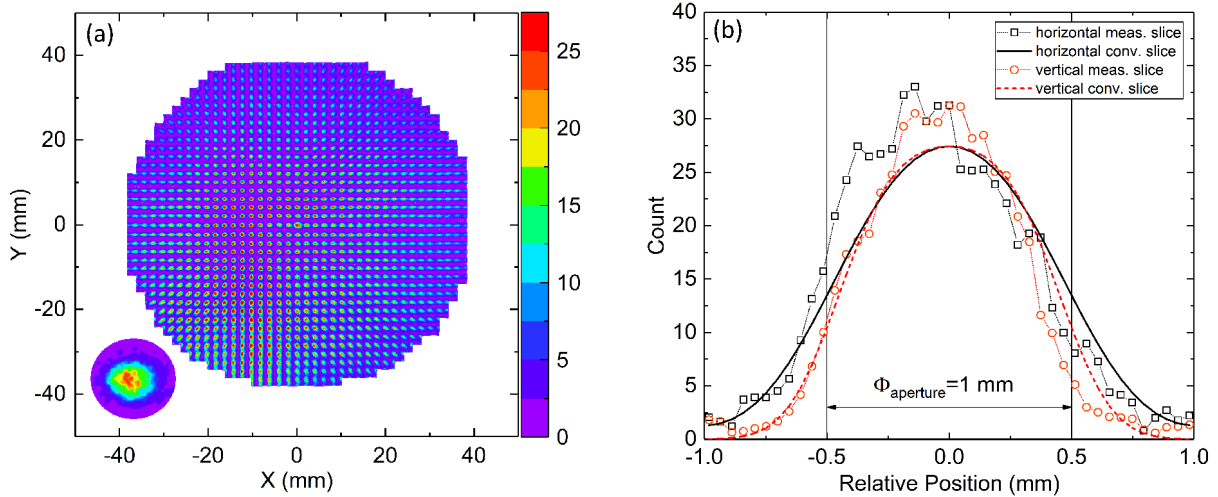

FIG. 8. (a) The position spectrum after calibration. The lower-left inset shows the aperture-spot at $x = -8$ mm and $y = 4$ mm. (b) Horizontal and vertical slices through the aperture center of the measured data and the two-dimensional convolution fitting results. The coordinate is relative to the aperture center for clarity.

## C. Calibration and resolution

The last two non-idealities make up the so-called nonlinearity effect as mentioned in section II. They can be corrected by calibration, in which the linearity as promised in the ideal case by the charge ratio equalities $Q_a^x/Q_b^x = R_b^x/R_a^x$ and $Q_a^y/Q_b^y = R_b^y/R_a^y$ is restored. The coordinates of the measured pixels is firstly shifted relative to the central aperture spot center. The row and column aperture spot centers are then fitted with polynomials (grid-polynomials), generating a nonlinear grid across the spectrum. Another pair of polynomials (intercept-polynomials) are further found, which fit the mapping from the grid-polynomial intercepts to the real-position space, taking into account the aperture optical shift effect. For a pixel in the channel spectrum, a couple of polynomials (pixel-polynomials) that "cross" at this pixel center are created by linear interpolation in the grid-polynomials' parameter space. The spatial coordinate of the current pixel is calculated with the intercept-polynomials. The current pixel's coverage in the position space is then determined in the same manner. Together with the pixel's count, the coverage directly gives the pixel's contributions to the discrete real-position space. In a word, treating the grid-polynomials as an interpolation scenario, the originally measured pixels are individually and cumulatively mapped to a discrete linear real-position space. The linear real-position space spans from -50 mm to 50 mm in both the X and Y directions, and also consists of 2048 × 2048 pixels. Notice that due to the fitting-interpolation nature, global rather than local linearity of the measured spectrum is guaranteed. The calibrated position spectrum is shown in Fig. 8(a), which corresponds to the effective detecting area indicated in Fig. 7c by the white polygon. Compared to the original spectrum, the spectrum asymmetry and nonlinearity are effectively corrected.

Taking the aperture spot at $x = -8$ mm and $y = 4$ mm for instance, the position resolution is further determined. The aperture spot is shown in the lower-left inset in Fig. 8(a). It is fitted with the convolution of the 2-dimensional ideal round window (aperture) function and a 2-dimensional Gaussian resolution function, taking the influence of the neighboring eight apertures into account. The horizontal and vertical slices through the aperture center of the measured data and the fitting result are shown in Fig. 8(b). The detector resolution is defined as the FWHM of the resolution function, which is fitted to be 0.62±0.01 mm in the horizontal direction and 0.38±0.01 mm in the vertical direction. Note that the influence of the optical expansion of the mask aperture and the effect of the finite radiation source area are treated as higher order infinitesimals and are ignored, considering the source-to-detector distance is much larger than the mask-to-MCP distance.

## V. CONCLUSION

We have developed a two-dimensional resistive anode, the DLSA anode, for a position sensitive MCP detector with a Ø100 mm sensitive area. The anode in the current configuration has a step response settling time of approximately 10 μs and advantages of easy manufacture, charge allocation robustness and flexible geometric applicability. The other main components of the detector are a Z-stack MCP configuration, a high-voltage circuit and four integrated spectroscopy amplifiers. The high voltage parameters of the detector are optimized to make sure that the last MCP works in a saturated state, and to balance the charge allocation ratio on the anode and the edge resolution deterioration of the detector. The spectroscopy amplifier is an improved version of the one in Ref.[4] with the parameters adapted to the current anode, which has a gain of 136.9 mV/pC, a full width of 20 μs, a linear correlation coefficient of 0.999994 and a baseline noise less than 10 mV. The detector is tested with an $^{55}$Fe x-ray source. The 1 mm apart Ø1 mm apertures in the mask are clearly distinguished. A global symmetry and linearity is guaranteed after calibration. The resolution of the current detector can reach 0.38 mm.

The detector has been applied in an MCT spectrometer. Future studies concerning the effects of the material, the thickness and the separation of the anode PCBs should be essential to improve the detector resolution, linearity and counting rate capability. For instance, copper cladding can be applied to the slot edges of the X anode to eliminate any uncertainties that might be caused by charging in higher counting rate situations, and the X anode can be made thinner to counteract the side effect of parasitic capacitance enhancement.

## ACKNOWLEDGEMENT

We gratefully acknowledge the technical assistance from the personnel in the Solid-State Detector Group, the Gas Detector Group, the EBIS Laboratory and the 320 kV Platforms for Multi-discipline Research with Highly Charged Ions at Institute of Modern Physics, Chinese Academy of Sciences. This work is supported by the National Natural Science Foundation of China under Grant Nos. 11774356 and 11675232.

[1] J. L.Wiza, Nucl. Instrum. Meth **162,** (1979).

[2] M. Lampton and F. Paresce, Review of Scientific Instruments **45,** 1098 (1974).

[3] C. Martin, P. Jelinsky, M. Lampton, R. F. Malina and H. O. Anger, Review of Scientific Instruments **52,** 1067 (1981).

[4] L. Yang, J. Liu, Y. Zhang, W. Wang, D. Yu, X. Li, X. Li, M. Zheng, B. Ding and X. Cai, Review of Scientific Instruments **88,** 086103 (2017).

[5] S. S. Gambhir, Nat. Rev. Cancer **2,** 683 (2002).

[6] K. Butter, P. H. H. Bomans, P. M. Frederik, G. J. Vroege and A. P. Philipse, Nat. Mater. **2,** 88 (2003).

[7] Y. Kaganovsky, D. Li, A. Holmgren, H. Jeon, K. P. MacCabe, D. G. Politte, J. A. O'Sullivan, L. Carin and D. J. Brady, J. Opt. Soc. Am. A **31,** 1369 (2014).

[8] G. W. Fraser and E. Mathieson, Nuclear Instruments and Methods **179,** 591 (1981).

[9] H. Spieler, *Semiconductor Detector Systems,* (Oxford University Press, New York, 2005), pp. 179.

[10] J. V. Hatfield, S. A. Burke, J. Comer, F. Currell, J. Goldfinch, T. A. York and P. J. Hicks, Review of Scientific Instruments **63,** 235 (1992).

[11] W. Wang, D. Yu, J. Liu, R. Lu and X. Cai, Review of Scientific Instruments **85,** 106104 (2014).

[12] C. Firmani, E. Ruiz, C. W. Carlson, M. Lampton and F. Paresce, Review of Scientific Instruments **53,** 570 (1982).

[13] J. Cazaux, Journal of Applied Physics **95,** 731 (2004).

[14] J. Cazaux, Scanning **26,** 181 (2004).

[15] B. Gross, G. M. Sessler and J. E. West, Journal of Applied Physics **45,** 2841 (1974).

[16] L. Nunes de Oliveira and B. Gross, Journal of Applied Physics **46,** 3132 (1975).